# A Comparative Analysis on LaueUtil and PRECOGNITION Software Packages as Tools in Treating the Small Molecule Time-Resolved Laue Diffraction Measurements at High Flux X-ray facilities

Authors

**Jose de Jesus Velazquez-Garcia[a]\*, Joanne Wong[b], Krishnayan Basuroy[a], Darina Storozhuk[a], Sofiane Saouane[c], Robert Henning[d] and Simone Techert[ae]\***

[a]Photon Science - Structural Dynamics in Chemical Systems, Deutsches Elektronen-Synchrotron DESY, Notkestraße 85, Hamburg, 22607, Germany

[b]Institut für Anorganische Chemie, Georg-August-Universität Göttingen, Tammannstraße 4, Göttingen, 37077, Germany

[c]PETRA III - P11 beamline, Deutsches Elektronen-Synchrotron DESY, Notkestraße 85, Hamburg, 22607, Germany

[d]Center for Advanced Radiation Sources, The University of Chicago, Argonne National Laboratory, 9700 South Cass Ave, Lemont, Illinois, 90439, USA

[e]Institut für Röntgenphysik, Georg-August-Universität Göttingen, Friedrich-Hund-Platz 1, Göttingen, 37077, Germany

Correspondence email: jose.velazquez@desy.de; simone.techert@desy.de

**Funding information**   The current work has been funded by the Deutsche Forschungsgemeinschaft (DFG, German Research Foundation) - 217133147/SFB 1073, projects B06, C02. HG-recruitment, HG-Innovation "ECRAPS", HG-Innovation DSF, DASHH, DGP and CMWS; BioCARS is supported by NIH grant P41 GM118217 and through a collaboration with P. Anfinrud (NIH/National Institute of Diabetes and Digestive and Kidney Diseases).

**Synopsis**   LaueUtil and PRECOGNITION software to treat time-resolved Laue crystallography data.

**Abstract**   Investigating metal organic systems with time-resolved photocrystallography poses a unique challenge while interpreting the time dependent photodifference maps. In these difference Fourier maps, the signals correspond to the movement of heavy metal atoms always overpower the signals from much lighter atoms attached to them. For a systematic assessment of the quality of the photodifference maps obtained from metal organic systems, in this work, LaueUtil and PRECOGNITION software were used to treat time-resolved Laue crystallography data of a [2x2] matrix-like Fe(II) complex. The rigid spot identification method in LaueUtil allows to identify and index > 250,000 reflections per 10 datasets. Though this leads to low completeness (< 30%), the software only treats the information from highly reliable diffraction spots. As a result, clean photodifference maps and small values in the thermal scale factor have been obtained. In the PRECOGNITION case, the package indexed more than 160,000 reflections per dataset. The resulting completeness is higher (>86%) and useful photodifference maps can be obtained by careful data treatment. However, the dependence on the refinement of the λ curve as well as the degradation of the sample may be reflected on the large thermal scale factors which also contribute as noise in the photodifference maps.

## 1. Introduction

The developments of synchrotrons and free-electron lasers have enabled the possibility of analysing short-living species in solid state with picosecond resolution by use of time-resolved (TR) crystallography methods (Bari et al. 2020). In such crystallography methods, a short laser pulse (pump) is rapidly followed by an interrogating X-ray pulse (probe) (Techert et al. 2001; Collet et al., 2003). Variation of the time delay between the pump and probe pulse gives information on the progress of the process subject to study. In stroboscopic experiments, the pump-probe cycle is repeated a large number of times to obtained sufficient counting statistics (Techert, 2004). Therefore, the crystal is exposed to a large number of repeating laser pulses which leads to rapid sample degradation. This is a serious limiting disadvantage for monochromatic methods where only small fraction of photons from the synchrotron beam is used and large exposure times are required. As a result, the pink Laue method ($\Delta E/E$ ~8%) has largely been used in TR crystallographic experiments due to the more efficient use of synchrotron photons and shorter exposure times (Coppens, 2011).

After data collection of TR Laue crystallographic experiment, the data processing of small molecules is usually done by different software such as LaueUtil (Kalinowski et al., 2011; Kalinowski et al., 2012) or PRECOGNITION (Ren, 2010). Both software kits are developed during the evolutionary stage of time-resolved X-ray crystallography, particularly at synchrotrons. The LaueUtil toolkit is based on the RATIO method (Coppens et al., 2009) in which the ratio of laser-ON and laser-OFF intensities ($R_{ON/OFF} = I_{ON}/I_{OFF}$) is used for the analysis instead of individual intensities. In this method for every single crystal orientation laser-OFF and laser-ON intensities are collected successively. Forming the ratio for every crystal orientation allows for a precise elimination of experimental long-term artefacts such as the dependence on the λ-curve for every orientation, or long-term drifts in the beamlines etc. As a disadvantage, LaueUtil is limited to time-resolved photocrystallographic experiments introducing small photo-induced structural changes from low photo-conversion percentages ($\leq 6\%$) and insignificant changes in the unit cell dimensions. A higher photo-conversion percentage > 6% or large changes in the unit cell dimension would cause a shift in the peaks position between the laser-OFF and laser-ON frames.

In contrast, PRECOGNITION separately treats laser-ON and laser-OFF datasets. This independent treatment gives the freedom of either collecting the laser-ON/laser-OFF data sets in a stroboscopic manner or collect the entire laser-ON or laser-OFF data sets separately (Messerschmidt et al., 2010). As a disadvantage, in PRECOGNITION, the reduction to a set of structure factors is affected by all the wavelength-dependent effects, such as absorption, anomalous scattering and the energy dependence of the detector response. This becomes a tedious work when high repeatability is sought but it is not so limited by variations in the unit cell dimensions.

In time-resolved studies of small molecules, it is common to choose one program above the others and then the research is limited by the availability and capabilities of the software. The utility of LaueUtil and PRECOGNITON have been demonstrated to their full capabilities when the cell's dimensions do not exhibit a significant change (Messerschmidt et al., 2010; Makal et al., 2012). However, it is important to determine the limits of the software if we pretend to have a full picture of the phenomena at different time scales. In this work, we briefly compare pink-Laue synchrotron data processed by LaueUtil and PRECOGNITION software on the time-resolved studies of a novel 2x2 iron (II) grid complex at -200 ps and 100 μs after light excitation. The -200 ps time point is in particular sensitive to noise fluctuations due to the experiment and data evaluation scheme, the 100 μs time point is a classical time point at which structural studies on heat effects (dissipated heat from heat waves) is performed.

The work gives an outline over currently existing LaueUtil and PRECOGNITION software toolkits which have been expanded (PRECOGNITION) and developed (LaueUtil) for time-resolved small molecule crystallography studies at large scale X-ray facilities. We compare the two software tools and discuss them not only in the context of experimental conditions but also with respect to the quality of the difference Fourier maps obtained for selected example.

**2. Materials and Methods**

**2.1. Materials**

The complex $[Fe_4L_4]4BF_3·C_2H_3N$, {L=4-methyl-3,5-bis{6-(2,2′-bipyridyl)}pyrazole)}, here called as **FE4** grid, presents a [2x2] grid configuration made by four perpendicular ligands strands and 4 iron atoms (Figure 1). In solid state, the material crystallized in the $C2/c$ space group with two crystallographically independent metal ions: Fe1 and Fe2. Synthesis procedure and characterisation have been reported somewhere else (Schneider et al., 2013).

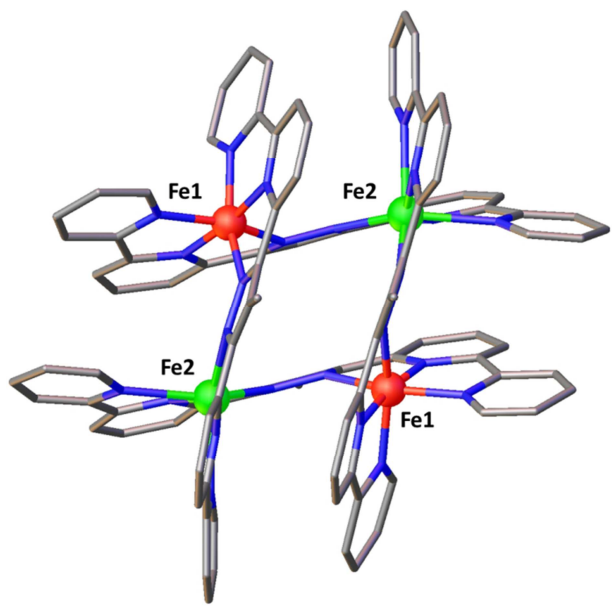

**Figure 1** Molecular structure of **FE4** grid, Fe1 atoms (red) and Fe2 atoms (green) are crystallographically independent metal ions.

### 2.2. Monochromatic Data Collection and Structural Information

A monochromatic dataset at 220 K was collected at the P11 beamline of the PETRA III at DESY (Hamburg, Germany). The collected dataset was used to determine the ground state structure without a thermal effect caused by repeated laser irradiation. The synchrotron data was collected at 20 keV and a ϕ rotation from 0 to 360° with a Δϕ value of 0.1°. XDS software (Kabsch, 1988 & 2010) was used for integration, scaling and space group assignment. The data was merged with the program SORTAV (Blessing, 1997). Finally, the structure solution and refinement was performed by SHELX (Sheldrick 2008 & 2015) combined with OLEX2 (Dolomanov et al., 2009) as GUI. Crystallographic information is listed in Table 1.

**Table 1** Crystallographic information of **FE4** at 220K.

| Space group | Z′ | $a$ (Å) | $b$ (Å) | $c$ (Å) | $\alpha$ (°) | $\beta$ (°) | $\gamma$ (°) |
| --- | --- | --- | --- | --- | --- | --- | --- |
| $C2/c$ | 0.5 | 30.919 (6) | 12.821 (3) | 27.159 (5) | 90 | 121.03 (3) | 90 |

### 2.3. Collection of Time-resolved Pink Laue Data

Time-resolved X-ray diffraction experiments have been performed at BioCARS beamline at the APS synchrotron (Argonne National Laboratory, Chicago, USA). The characteristics of the diffraction experiments can be seen in Table 1. The TR Laue diffraction data was collected at 220 K with a pink beam at 15 keV and a ϕ rotation from 0 to 180° with a Δϕ value of 1°. We used typical excitation densities of 2 mJ/mm$^2$ and -200 ps and 100 μs delay times between the laser pump and the X-ray probe. Long-range fluctuations in the beam's position or intensity are minimized by successive collection of laser-OFF and laser-ON frames. Such OFF/ON cycle was repeated five times for each frame.

**Table 2** Experimental characteristics of the time-resolved experiments.

| Characteristic | Crystal average dimension | Temperature | Laser power | Optical excitation wavelength | Rep. rate | φ rotation | Delay times | |
|---|---|---|---|---|---|---|---|---|
| Value | ≈100 | 220 | 2 | 390 | 41.1 | 0-180 | -200 | 100 |
| Units | μm | K | mJ | nm | Hz | ° | ps | μs |

Figure 2 gives a schematic representation of the three types of data treated in this work. In the notation of the states studied in this work we will not use Schoenflies notation but description of the states which refer for clarity to the practical experiment. Here, we will refer as "Mono" to the monochromatic data collected at 220 K completely without laser excitation. "laser-OFF" and "laser-ON" data refer to the data collected during the "laser-OFF" and "laser-ON" cycle in the stroboscopic experiment. Since we concentrate in this paper on exact refinement of the 100 μs state, our "laser-ON" is equal to transient 100 μs state.

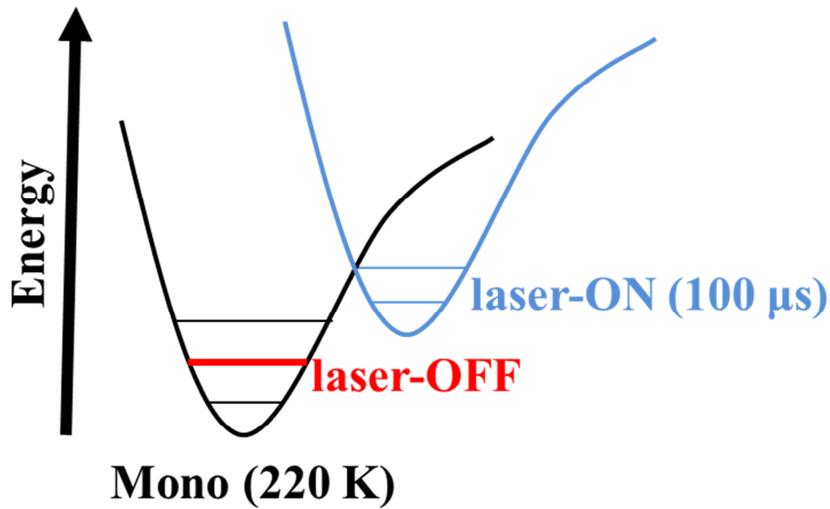

**Figure 2** Schematic representation of "Mono", "laser-OFF" and "laser-ON" energy levels.

An additional dataset at -200 ps delay time was taken in order to analyse the remaining heat in the sample. Here, the probe pulse interacts with the sample 200 ps before the pump pulse. In this way, any remaining heat inside of the sample or a long-lasting excited state will be detected in the temperature analysis (section 3.2) or in the photodifference maps (section 3.3).

**2.4. Data Processing**

The unit cell dimensions obtained from the monochromatic experiment have been used as an input for the PRECOGNITION software. The results reported here are based on linear analytical integration for one pair of laser-OFF and laser-ON dataset for each delay time: -200 ps and 100 µs. In time-resolved studies, it is crucial to bring laser-OFF and laser-ON data sets in the same scale. In PRECOGNITION, an individual program Epinorm (energy partial improved normalization) is used for data scaling and merging. Epinorm can scale together the laser-OFF and laser-ON data sets and separately merge them. However, we use Epinorm to separately scale laser-ON and laser-OFF data and validate the reconstructed source spectrum. The obtained reflection files were normalized and standardized using a python script developed in the group. Once the structure factor amplitudes are obtained, the repeatedly measured and symmetry-related reflections were averaged by the program SORTAV. Finally, the structure of the laser-OFF and laser-ON are solved and refined by SHELX and OLEX2 as GUI.

The intensities from the full laser OFF/ON cycle were integrated and indexed by LaueUtil. The integration is based on the statistical analysis of the values from each of the pixels on successive frames which does not require previous knowledge of the sample's cell dimensions. However, the orientation matrix determination method, although fast, requires the knowledge of the unit cell parameters and approximate structure factors correspond to the diffraction spots, obtained from the monochromatic data. Every diffraction spot is assigned to a particular hkl value. After the statistical analysis of 5 laser OFF/ON frames, ratio values between laser-ON and laser-OFF were calculated. The obtained ratios were averaged with the program SORTAV.

Once structure factor amplitudes from PRECOGNITION and ratios from LaueUtil are calculated, photodifference maps are generated by Olex2. The photodifference maps share the formula (Coppens et al., 2008; Makal et al., 2011):

$$\Delta\rho(\boldsymbol{r}) = \rho^{laserON}(\boldsymbol{r}) - \rho^{laserOFF}(\boldsymbol{r}) =$$

$$\left(\frac{1}{V}\right)\sum[F_{obs}^{laserON}(\boldsymbol{H}) - F_{obs}^{laserOFF}(\boldsymbol{H})]\exp[i\varphi_{calc}^{laserOFF}(\boldsymbol{H})]\exp[-i2\pi\boldsymbol{H}\cdot\boldsymbol{r}] \qquad (1)$$

In the case of PRECOGNITION, equation 1 is used for all the photodifference maps. In the case of the RATIO method used in LaueUtil, a "semi-observed" laser-ON structure factor is defined as (Coppens & Fournier, 2015):

$$F_{semi-obs}^{laserON}(\boldsymbol{H}) = [R_{obs}(\boldsymbol{H})^{1/2}\, F_{calc}^{Mono}(\boldsymbol{H})] \qquad (2)$$

Therefore, photodifference maps obtained from LaueUtil are calculated by:

$$\Delta\rho(\boldsymbol{r}) = \left(\frac{1}{V}\right)\sum[R_{obs}(\boldsymbol{H})^{1/2} - 1]\; F_{calc}^{Mono}(\boldsymbol{H})\exp[i\varphi_{calc}^{laserOFF}(\boldsymbol{H})]\exp[-i2\pi\boldsymbol{H}\cdot\boldsymbol{r}] \qquad (3)$$

## 3. Results and Discussion

### 3.1. Characteristic of the Time-resolved Datasets

The characteristics of datasets evaluated by both software packages are shown in Table 3. For the LaueUtil data, the software indexed more than 250,000 reflections per 10 datasets which accounts for 85% of the detected reflections. The indexed reflections are used to calculate $R_{ON/OFF}$ values and subsequently merged datasets for different $\sigma(R)$ cut-offs. The merged datasets count between 1725 and 2556 unique reflections for completeness between 21.3% and 36.7% for both delay times.

In the PRECOGNITON case, the software indexed more than 160,000 reflections per dataset. Merged datasets were generated by separately merging laser-OFF and laser-ON reflections with a cut-off $\sigma(I) \leq 0.5$. The generated datasets contain 8132 and 8131 unique reflections and completeness of 86.7% and 87.4% for laser-ON and laser-OFF datasets at 100 μs, respectively.

**Table 3** Characteristics of the evaluated datasets by LaueUtil.

| Software | dataset | Delay time | $\sigma^a$ | No of datasets | Indexed reflections | Number of unique reflections | Completeness (%) | R1 Factor |
|---|---|---|---|---|---|---|---|---|
| LaueUtil | ON/OFF | 100 μs | 0.1 | 10 | 338482 | 1725 | 27.5 | --- |
|  | ON/OFF | 100 μs | 0.5 | 10 | 338482 | 2556 | 36.1 | --- |
|  | ON/OFF | 100 μs | 0.7 | 10 | 338482 | 2596 | 36.7 | --- |
|  | ON/OFF | -200 ps | 0.1 | 10 | 267442 | 1561 | 21.3 | ----- |
| ---------- | ---------- | -------- | -------- | ---------- | ------------ | ---------- | --------------- | -------- |
| PRECOG-NITION | OFF | 100 μs | 0.5 | 1 | 164103 | 8132 | 87.4 | 0.211 |
|  | ON | 100 μs | 0.5 | 1 | 163904 | 8131 | 86.7 | 0.1782 |
|  | ON | -200 ps | 0.5 | 1 | 123649 | 8191 | 89.9 | 0.1827 |

[a]Values under the σ column corresponds to σ(R) and σ(I) for LaueUtil and PRECOGNITION data, respectively.

For LaueUtil, the spot detection was done by the seed-skewness method (Bolotovsky et al., 1995; Szarejko et al., 2020) which does not require a predicted spot position and proved to be reliable for spots with weak and medium intensities. The procedure of spot identification was performed over the five laser-ON and five laser-OFF frames. If a spot does not get detected in all the 10 frames in a block with intensity that satisfies the given cut-off then that spot would be rejected. As a result, we move forward in the data processing with much more reliable spots for LaueUtil. This can be also checked if one sees the number of spots indexed by LaueUtil/PRECOGNITION and the number of unique ratios/reflections values presented in Table 3. In order to have the same level of completeness as PRECOGNITION, one can relaxes the spot identification step in LaueUtil and allowed more spots to get identified with the risk of introducing more noise into the data.

### 3.2. Analysis of the Crystal's Temperatures

An estimate of the temperature increase due to laser irradiation can be obtained from the Wilson plots for the PRECOGNITION data and from the photo-Wilson plots for the LaueUtil data (Figure 3). The slope of the Wilson (equation 4) (Wilson, 1942) and photo-Wilson plot (equation 5) (Schmøkel et al., 2010) corresponds to twice the Debye-Waller factor (B) and twice the increase of B-factor (ΔB), respectively:

$$\ln\left(\frac{<I^2>}{\Sigma f^2}\right) = -2B\left(\frac{\sin\theta}{\lambda}\right)^2 + c \qquad (4)$$

$$\ln\left(R_{\frac{ON}{OFF}}\right) = -2\Delta B\left(\frac{\sin\theta}{\lambda}\right)^2 + b \qquad (5)$$

From the values of B and ΔB, the temperature scale factor ($k_B$) can be obtained by (Ozawa et al., 1998):

$$k_B = \frac{(B^{OFF}+\Delta B)}{B^{OFF}} = 1 + \frac{\Delta B}{B^{OFF}} \qquad (6)$$

In the case of LaueUtil, ΔB as a difference and $k_B$ as a ratio are only reported for $\Delta B^{laserON-laserOFF}$ and $k_B^{laserON/laserOFF}$. From PRECOGNITION we can calculate not only $\Delta B^{laserON-laserOFF}$ / $k_B^{laserON/laserOFF}$ but also $\Delta B^{laserOFF-Mono}$ / $k_B^{laserOFF/Mono}$ and $\Delta B^{laserON-Mono}$ / $k_B^{laserON/Mono}$. In this way the values of $k_B^{laserON/laserOFF}$ accounts for a temperature increase during the laser-ON period while $k_B^{laserOFF/Mono}$ provides an idea of any temperature change due to accumulated heat during the data collection. In the case of no heating $k_B^{laserOFF/Mono} = 1$ and $\Delta B^{laserOFF-Mono} = 0$. Finally, $k_B^{laserON/Mono}$ indicates the relative temperature increase.

The values of ΔB and $k_B$ are reported in Table 4. In the LaueUtil case, the temperature scale factor at 100 μs indicates a minimal temperature increase between the laser-ON and the laser-OFF data. However, the values of ΔB are comparable with those from other monometallic iron complexes at the same delay time (Lorenc et al., 2009). Additionally, the smaller values of $k_B$ and ΔB at -200 ps indicates that there is no heat accumulated during the stroboscopic experiment. Contrary, the values of $k_B$ obtained from PRECOGNITION data at 100 μs suggest a considerable temperature increase not only between the laser-ON and laser-OFF but also between the Mono and the laser-OFF data. Meanwhile the $k_B^{laserON/laserOFF} > 1$ indicates a rise in temperature caused by the laser irradiation; the $k_B^{laserOFF/Mono} >1$ implies that the system is heated in the excited or ground state during the whole data collection. This statement is also supported by the large values of the temperature scale factor at -200 ps. Additionally, the fact that $k_B^{laserOFF/Mono} > k_B^{laserON/OFF}$ indicates that the temperature rise between the laser-OFF and Mono is considerably bigger than that between the laser-OFF and the laser-ON.

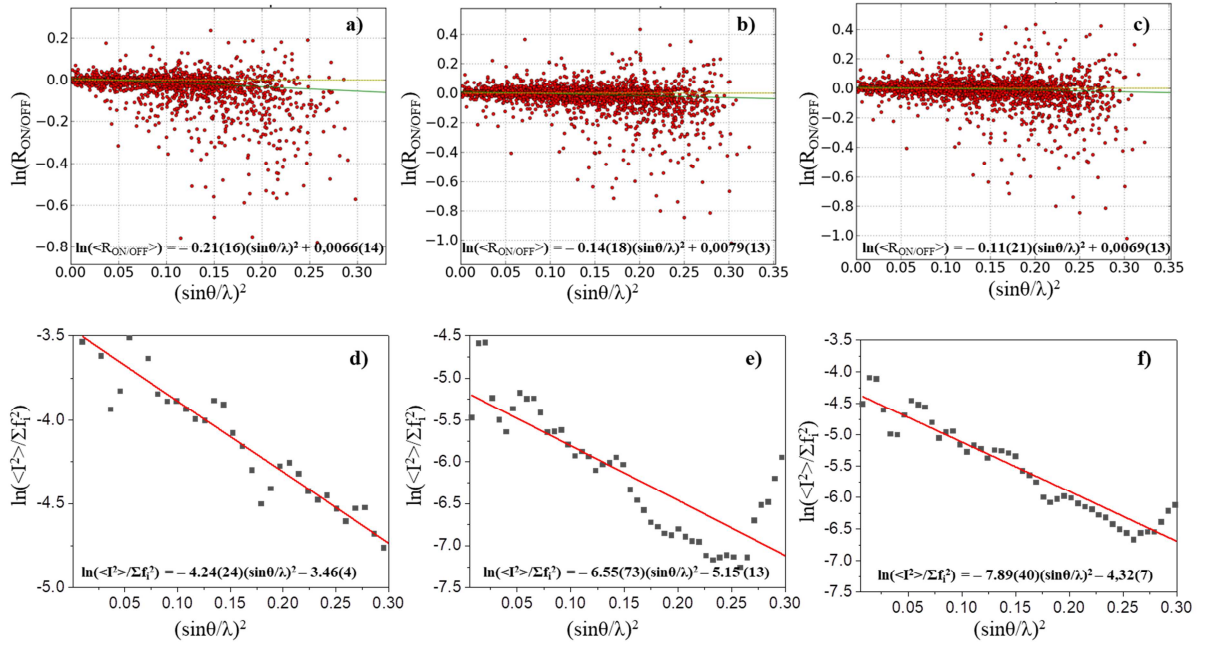

**Figure 3** Photo-Wilson plot at 100 ps for LaueUtil treatment with a-c) cut-off $\sigma(R) \leq 0.1$, 0.5 and 0.7, respectively. d) Wilson plot for Mono data and e-f) laser-OFF and laser-ON data at 100 μs treated with the PRECOGNITION software.

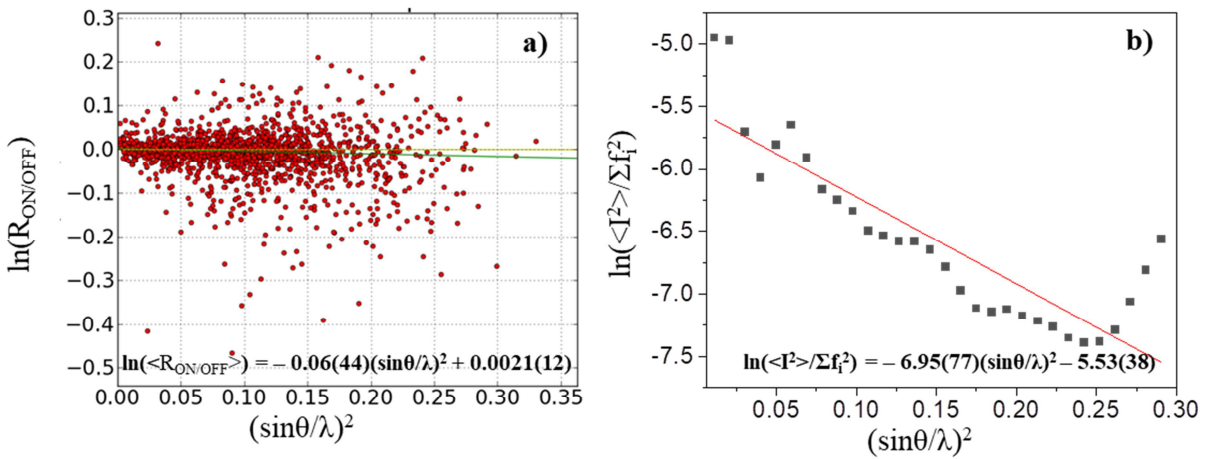

**Figure 4** a) Photo-Wilson plot of the -200 ps, LaueUtil treated data and b) Wilson plot of the – 200 ps PRECOGNITION treated data, respectively.

**Table 4** Comparison of ΔB and $k_B$ values for all datasets.

|  |  | delay time | σ[b] | ΔB (Å$^2$) | $k_B$ |
|---|---|---|---|---|---|
| LaueUtil | ON/OFF | 100 μs | σ(R) ≤ 0.1 | 0.106 (77) | 1.0066 (15) |
|  | ON/OFF | 100 μs | σ(R) ≤ 0.5 | 0.068 (87) | 1.0080 (13) |
|  | ON/OFF | 100 μs | σ(R) ≤ 0.7 | 0.056 (104) | 1.0070 (13) |
|  | ON/OFF | -200 ps | σ(R) ≤ 0.1 | 0.03 (22) | 1.0021 (11) |
| PRECOG. | OFF/Mono[a] | 100 μs | σ(I) ≤ 0.5 | 1.16 (77) | 1.54 (27) |
|  | ON/Mono[a] | 100 μs | σ(I) ≤ 0.5 | 1.83 (47) | 1.86 (4) |
|  | ON/OFF | 100 μs | σ(I) ≤ 0.5 | 0.67 (83) | 1.21 (21) |
|  | ON/Mono[a] | -200 ps | σ(I) ≤ 0.5 | 1.35 (80) | 1.64 (28) |

[a] In the PRECOGNITION datasets, values of ΔB and $k_B$ corresponding to OFF/Mono refers to ΔB$^{laserOFF-Mono}$ and $k_B^{laserOFF-Mono}$, meanwhile ON/Mono refers to ΔB$^{laserON-Mono}$ and $k_B^{laserON/Mono}$. [b] Values under the σ column corresponds to σ(R) and σ(I) for LaueUtil and PRECOGNITION data, respectively.

### 3.3. Photodifference Maps

The photodifference maps based on all independent reflections from the dataset processed by PRECOGNITION and LaueUtil are shown in Figure 5. Isosurfaces are drawn at 0.13 eÅ$^{-3}$ and 0.27 eÅ$^{-3}$ for LaueUtil and 1.6 eÅ$^{-3}$ and 4.5 eÅ$^{-3}$ for PRECOGNITION data correspond to -200 ps and 100 μs delay times, respectively. In both cases, there is not a clear indication of displacement towards a particular direction from any of the crystallographically independent metal atoms. Any molecular response upon excitation at 100 μs is largely overtaken by the effects caused by heatwaves and diffuse heating processes. The presence of the diffuse heating is corroborated by the large values of the thermal scale factor and ΔB reported in Figure 4 and section 3.2.

A considerable difference in result between methods is seen at negative delay times. In the LaueUtil case, the empty photodifference maps and low thermal scale values at -200 ps indicates that the system returns to its ground state after laser excitation and it is not heated in the excited or ground state during the stroboscopic experiment and the way difference maps are collected and indexed (laser-ON / laser-OFF collections for single crystal orientations). While evaluating the data with PRECOGNITION, the non-vanishing densities in the photodifference maps seem to suggest the "opposite": The residual photodifference maps as well as the large value of the thermal scale factors suggest that the system is heated during data collection. Since the data evaluation is based on the same data set and since the crystal is surely not heated when treated with different software, the differences at negative time points reflect experimental boundary conditions which need to be fulfilled when the PRECOGNITION software is been used. The recommendations in experimental conditions for both programs are provided in section 3.4.

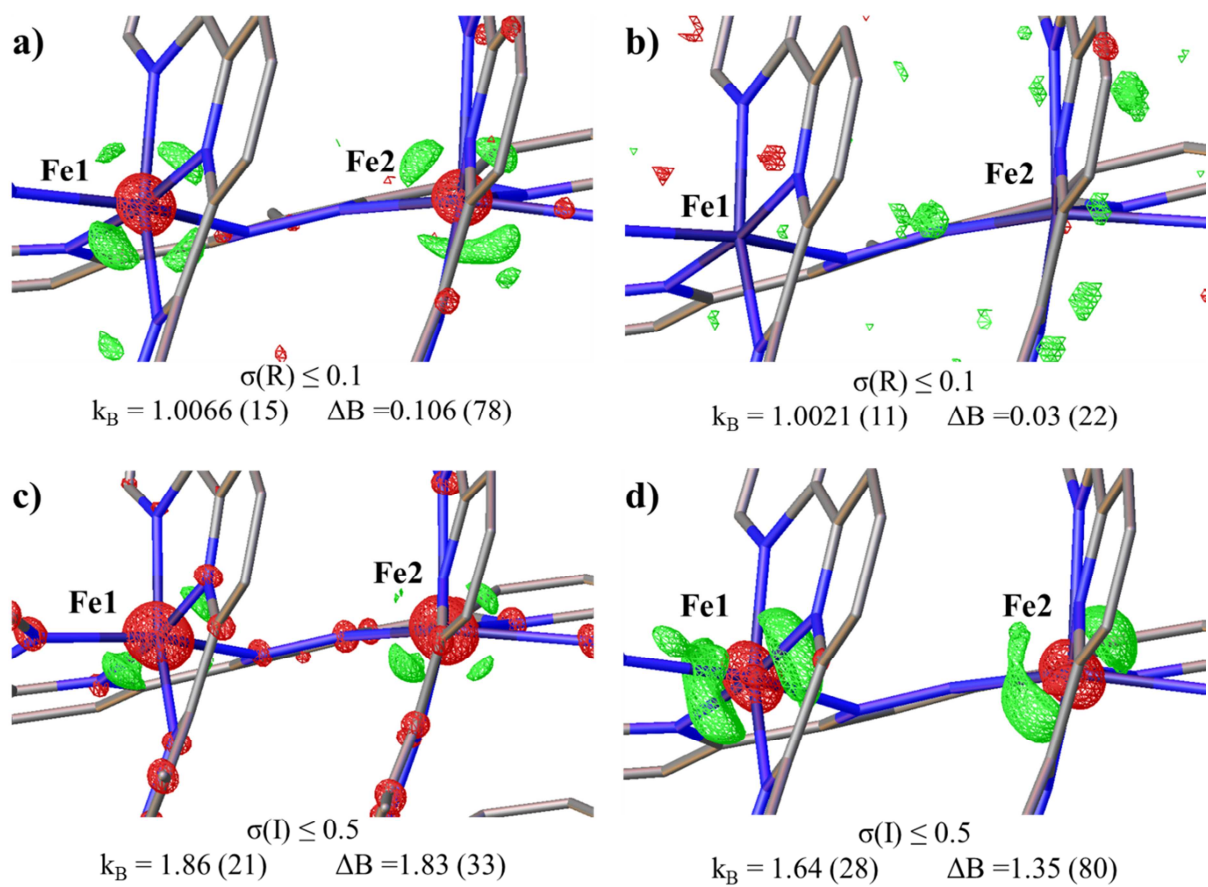

**Figure 5** Photodifference maps generated from LaueUtil and PRECOGNITION for the same sets of data. a) LaueUtil electron density difference map of the (100us - OFF) data, cut-off σ(R) ≤ 0.1; b) LaueUtil electron density difference map of the (-200ps - OFF) data, cut-off σ(R) ≤ 0.1; c) PRECOGNITION electron density difference map of the (100us - OFF) data, cut-off σ(I) ≤ 0.5; d) PRECOGNITION electron density difference map of the (-200ps - OFF) data, cut-off σ(I) ≤ 0.5. Isosurfaces are drawn at 0.13 eÅ$^{-3}$ and 0.27 eÅ$^{-3}$ for LaueUtil, and 1.6 eÅ$^{-3}$ and 4.5 eÅ$^{-3}$ PRECOGNITION data for the -200 ps and 100 μs delay times, respectively. Fe1, Fe2 and neighbouring ligand atoms are only focussed.

### 3.4. Analysis

The basic steps to treat time-resolved photocrystallography data with LaueUtil and PRECOGNITION are summarized in Fig. 6. The chart exhibits the clear differences in the steps to treat time-resolved XRD data which are also a reflection of the differences between methods used by the two software tools. PRECOGNITION treats the Laue diffraction data individually and on an absolute scale by separately evaluating laser-OFF and laser-ON data intensities, sorted and merged to give complete laser-OFF and laser-ON data sets. As a result, it provides structure factor amplitudes from which the final laser-OFF and laser-ON structures are obtained.

LaueUtil is based on the RATIO method and reduces the information to a set of ratios between the laser-ON and laser-OFF intensities for every spot and then sorted and merged over the whole Ewald sphere. From the calculated ratios, it is not possible to obtain a refined structure applying further conventional software such as SHELX or Olex2. Secondary software (LASER2010, Vorontsov et al., 2010) is required to separate the refinement of the transient light-induced structural changes, corresponding thermal scale factors and population of the transient species, which are not reported in the present publication (here we report only initial temperature scale factor and photodifference maps from the calculated ratios).

As a result of these different approaches, PRECOGNITION yields in higher completeness of the data and allows for the refinement of bigger structural changes, while the approach of LaueUtil limits to a smaller completeness with only the relevant yielding in less noisy photodifference maps. Finally, the differences in the methodologies lead to the reported discrepancies in the photodifference maps and thermal scale factors. Intrinsically $k_B$ is higher in the data process by PRECOGNITION than LaueUtil, due to the increased uncertainty in the refinement of the lambda curve for both laser-OFF and laser-ON data sets and due to small experimental fluctuations and sample degradation. On one hand, LaueUtil avoids such a problem by using the RATIO method and the seed-skewness approach for spot identification. However, LaueUtil only provides the prior-refinement photodifference maps and thermal scale factors while those from PRECOGNITION are calculated from the refined structures from the laser-ON and laser-OFF data. On the other hand, PRECOGNITION is more sensitive to the crystal quality and its degradation: non vanishing Fourier difference maps at negative time-points allow deriving realistic experimental boundary conditions about disorder of the photo excited system which needs to be avoided when the PRECOGNITION software is applied.

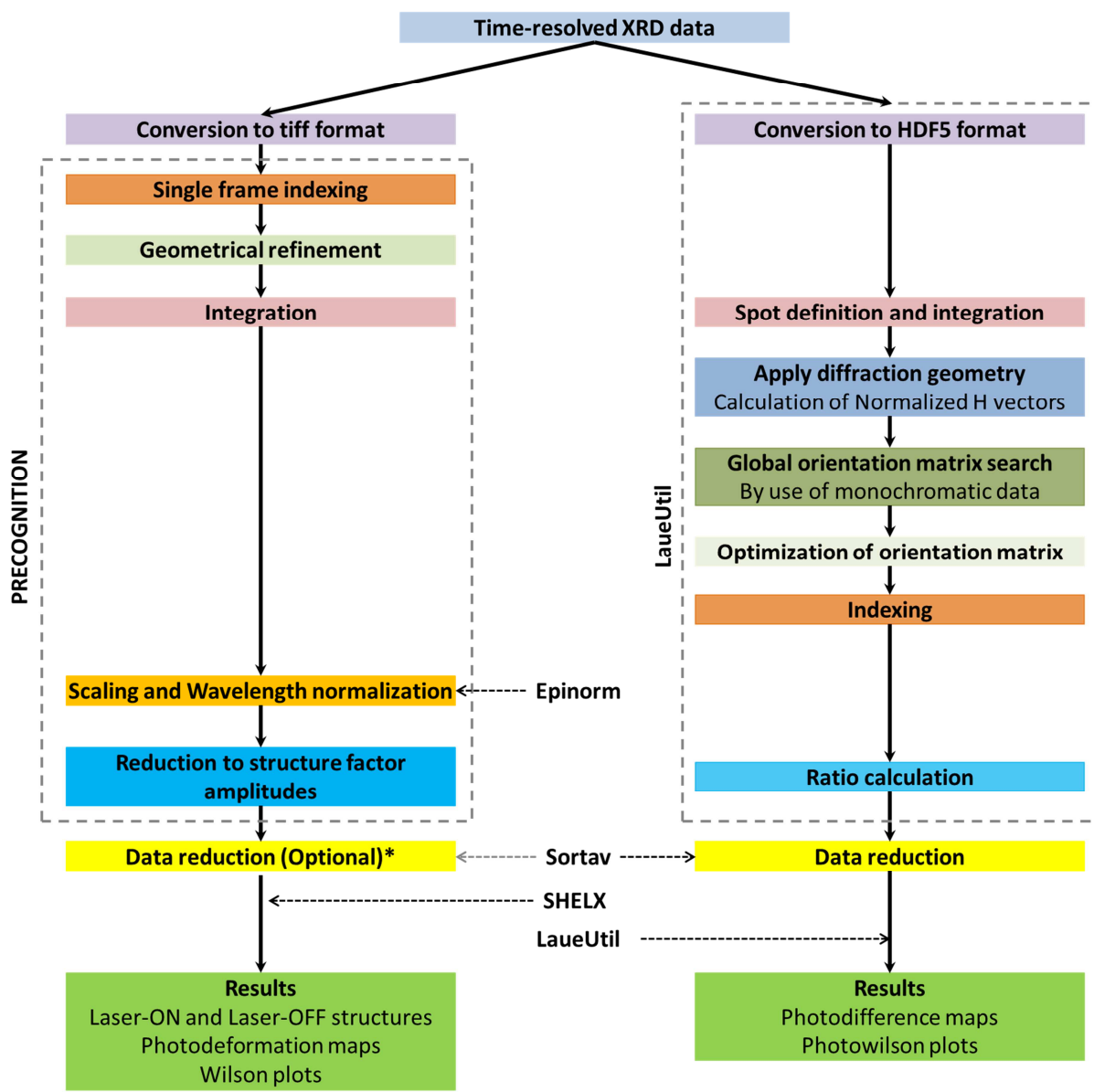

**Figure 6** Comparison of elementary steps for treating small molecule time-resolved data with PRECOGNITION and with LaueUtil. The right side has been adopted from reference Coppens et al., 2015. *Sortav package is not necessary when the data is scaled and merged by Epinorm, see section 2.4.

Taking into account the differences in methodology between packages, we have the following recommendations while collecting and treating the time-resolved data of small molecules:

a)  PRECOGNITION:

    a. This program is recommended for samples with large photo-induced structural changes or photo-conversions that can produce large changes in the unit cell dimensions.

b. We suggest a data collection strategy of one laser-OFF, one laser-ON for samples with low stability under light irradiation. A stroboscopic experiment of 5 laser-OFF, 5 laser-ON or more may lead to large temperature scale factors and introduce noise to the photodifference maps. Refer to appendix A for examples of structures solved with PRCOGNITION and the recommended data collection strategy.

c. When larger data completeness or better statistics is required for one delay time, different samples must be used.

b) LaueUtil:

a. We recommend the program for samples that requires large statistic to visualise small structural changes or changes caused by the low photo-conversion.

b. A stroboscopic experiment of 10 laser-OFF, 10 laser-ON is recommended for crystals stable under light irradiation. While for samples with low stability under light irradiation, it is recommended to have a cycle of 5 laser-OFF, 5 laser-ON.

## 4. Conclusion

In this work LaueUtil and PRECOGNITON have powerfully demonstrated their utility to treat time-resolved crystallography data of small molecules at high flux X-ray facilities. By comparing the two software tools and discussing them not only in the context of experimental conditions but also with respect to the quality of the photodifference maps of the selected example, we have provided a guideline for a) setting up the optimal experimental conditions for a time-resolved small molecule photocrystallography measurements and b) time-resolved studies correspond to which system benefits most from which type of applied software toolkit.

While investigating a crystalline **FE4** grid system with expected small positional changes of the atomic positions upon photo-excitation, we have treated the same time-resolved crystallography data sets and evaluated the structural changes for the -200 ps and 100 µs time points. The outcome obtained by both the software packages and including the results of previous studies suggests that PRECOGNITION allows for indexing to very high completeness of data sets. Due to the independent refinement of laser-ON and laser-OFF structural parameters within the Ewald sphere the indexed data are very robust against high photo-excitation yields and big structural changes. PRECOGNITION is therefore of high value for investigating large structural changes in molecular systems. As a disadvantage, the results from PRECOGNITION are largely affected by their dependence on the λ-curve and experimental ambiguities leading intrinsically to large thermal scale factors which add to the noise of the experimental difference Fourier maps potentially smearing out time-dependent structural changes and their amplitudes. In particular small structural changes and atomic displacements might therefore be difficult to be observed when PRECOGNITION is used for data evaluation.

Contrary, LaueUtil avoids such problem by applying the RATIO method. The results show higher relevance, in particular when small molecular geometric changes are expected, due to the "filtering in collection mode" which reduces the noise in the difference Fourier maps. As a disadvantage, the extended processing schemes and secondary software are required to refine the geometry and thermal structure factors. Furthermore, the software limits to systems with significant changes in the unit cell parameters.

In summary we have explored the importance of both the software tools and their complementarity for tackling time-resolved diffraction data of small molecules. Depending on the scientific questions, a prior selection of software and data collection strategy is required for high flux X-ray beamline. A small guidance of a successful approach has been worked out in this work.

**Acknowledgements** This research used resources of the Advanced Photon Source, a U.S. Department of Energy (DOE) Office of Science User Facility operated for the DOE Office of Science by Argonne National Laboratory under Contract No. DE-AC02-06CH11357. Use of BioCARS was also supported by the National Institute of General Medical Sciences of the National Institutes of Health under grant number P41 GM118217. Time-resolved set-up at Sector 14 was funded in part through a collaboration with Philip Anfinrud (NIH/NIDDK). The content is solely the responsibility of the authors and does not necessarily represent the official views of the National Institutes of Health. Portions of this research were carried out at the light source PETRA-III at DESY, a member of the Helmholtz Association (HGF). We would like to thank P11 staff for assistance in using beamline P11 and C. Paulmann and M. Tolkiehn for assistance in using beamline P24. We are grateful for support and suggestion by Bertrand Fournier on the use of LaueUtil and Vukica Šrajer on the use of PRECOGNITION.

**References**

Bari, S. Thekku Veedi, S. & Techert**,** S. in Salditt T. et al. (eds.), *Nanoscale Photonic Imaging*, *Springer Series: Topics in Applied Physics* **134** (2020).

Blessing, R. H. (1997) *J. Appl. Cryst*. **30**, 421-426.

Bolotovsky, R., White, M. A., Darovsky, A. & Coppens, P. H. (1995). *J. Appl. Cryst*. **28**, 86-95.

Collet, E., Lemée-Cailleau, M.-H., Buron-Le Cointe, M., Cailleau, H., Wulff, M., Luty, T., Koshihara, S.-Y. Meyer, M., Toupet, L., Rabiller, P. & Techert, S. (2003) *Science*. **300**, 612-615

Coppens, P. (2011). *J. Phys. Chem. Lett*. **2**, 616–621

Coppens, P. & Fournier, B. (2015) *J. Synchrotron Rad*. **22**, 280–287.

Coppens, P., Zheng, S.-L. & Gembicky, M. (2008). *Z. Kristallogr*. **223**, 265–271.

Coppens, P., Pitak, M., Gembicky, M., Messerschmidt, M., Scheins, S., Benedict, J., Adachi, S., Sato, T., Nozawa, S., Ichiyanagi, K., Chollet, M. & Koshihara, S. (2009) *J. Synchrotron Rad*. **16**, 226–230.

Dolomanov, O. V., Bourhis, L. J., Gildea, R. J., Howard, J. A. K. & Puschmann, H. (2009). *J. Appl. Cryst.* **42**, 339–341.

Kabsch, W. (1988) *J. Appl. Cryst.* **21**, 916–924.

Kabsch, W. (2010) *Acta Cryst.* D**66**, 133–144.

Kalinowski, J. A., Makal, A. & Coppens P. (2011) *J. Appl. Cryst.* **44**, 1182-1189.

Kalinowski, J. A., Fournier, B., Makal, A. & Coppens P (2012) *J. Synchrotron Rad.* **19**, 637-646.

Makal, A., Benedict, J., Trzop, E., Sokolow, J., Fournier, B., Chen, Y., Kalinowski, J. A., Graber, T., Henning, R. & Coppens, P. (2012*) J. Phys. Chem.* A**116**, 3359–3365.

Makal, A., Trzop, E., Sokolow, J., Kalinowski, J., Benedict, J. & Coppens, P. (2011). *Acta Cryst.* A**67**, 319–326.

Messerschmidt, M., Busse, G., Davaasambuu, J., Camaratta, M., Meens, A., Tschentscher, T. & Techert, S. (2010) *J. Phys. Chem.* A **114**, 7677-7681.

Lorenc, M., Hebert, J., Moisan, N, Trzop, E., Servol, M., Buron-Le Cointe, M., Cailleau, H., Boillot, M. L., Pontecorvo, E., Wulff, M., Koshihara, S., & Collet, E. (2009) *Phys. Rev. Lett.* **103**, 028301-1-4

Ozawa, Y., Pressprich, M. R. & Coppens, P. (1998) *J. Appl. Cryst.* **31**, 128-135.

Ren, Z. (2010) PRECOGNITION User Guide. Renz Research Inc., Illinois, USA.

Schneider, B., Demeshko, S., Neudeck, S., Dechert, S. & Meyer, F. (2013) *Inorg. Chem.* **52**, 13230-13237.

Sheldrick, G.M. (2008) *Acta Cryst.* A**64**, 112-122.

Sheldrick, G.M. (2015) *Acta Cryst.* C**71**, 3-8.

Schmøkel, M., Kaminski, R., Benedict, J. B. & Coppens, P. (2010) *Acta Crystallogr.* A**66**, 632-66

Szarejko, D., Kamiński, R., Łaski, P., & Jarzembska, K. N. (2020) *J. Synchrotron Rad.* **27**, 405-413

Techert, S., Schotte, F., Wulff, M. (2001) *Phys. Rev. Lett.* **86**, 2030-2034.

Techert, S. (2004) *J. Appl. Cryst.* **37**, 445-458.

Wilson, A. J. C. (1942) *Nature*. **150**, 152.

Vorontsov, I., Pillet, S., Kamiński, R., Schmøkel, M. S. & Coppens, P. (2010). *J. Appl. Cryst.* **43**, 1129–1130.

**Appendix A. Related structures**

Time-resolved structures related to this research have been submitted to CCDC, deposition numbers: 1942203-1942235, 1942291-1942299, 1942303-1942308, 1942312-1942320, 1942327-1942332, 1942334-1942342.